\documentclass[twocolumn]{article}

\usepackage{cite}
\usepackage{amsmath}
\usepackage{amsfonts}
\usepackage{algorithmic}
\usepackage{array}
\usepackage{hyperref}
\usepackage{cancel}
\usepackage{xcolor}        

\usepackage{graphicx}
\usepackage{url}

\newcommand{\myurl}[1]{\textcolor{blue}{\url{#1}}}
\newcommand{\myhref}[2]{\textcolor{blue}{\href{#1}{#2}}}

\newcommand{\refmmlist}[2]{\myhref{https://www.cantab.net/users/mmlist/#1}{#2}}
\newcommand{\refclass}[3]{\refmmlist{MML/A/doc/#1/#2.html}{#3}}
\newcommand{\refmember}[4]{\refmmlist{MML/A/doc/#1/#2.html\##3}{#4}}

\newcommand{\ignore}[1]{}
\newcommand{\mysec}[1]{section \ref{#1}}
\newcommand{\myfig}[1]{figure \ref{#1}}

\newcommand{\myMathit}[1]{\text{\textit{#1}}}     
\newcommand{\pr}[0]{\mathrm{pr}}

\begin{document}


\title{Subclasses of Class Function used to
       Implement Transformations of Statistical Models}

\author{Lloyd Allison, \\
  Faculty of Information Technology, \\
  Monash University, Clayton, Victoria 3800, Australia \\
  lloyd.allison@monash.edu \\
}

%

\date{}

\maketitle

\begin{abstract}
A library of software for inductive inference
guided by the Minimum Message Length (MML) principle was created previously.
It contains various (object-oriented-) classes and subclasses of
statistical Model and can be used to infer Models from given data sets
in machine learning problems.
Here transformations of statistical Models are considered and
implemented within the library so as to have desirable properties from
the object-oriented programming and mathematical points of view.
The subclasses of class Function needed to do such transformations are defined.
\end{abstract}

\textbf{keywords:}
  Statistical Model, transformation, class Function,
  machine learning, inference, information, MML


\maketitle


\section{Introduction}
\label{sec:Intro}

A library of
software\footnote{See
   \myurl{https://www.cantab.net/users/mmlist/MML/A/}
   for source-code and documentation.}
based on the Minimum Message Length (MML) principle \cite{WB68,Wal05}
was created previously \cite{All18, mmlist}
for use in writing programs to solve machine learning problems,
that is to infer statistical Models from given data sets.
It follows on from an earlier prototype\cite{All05}.
The software defines various
classes\footnote{In the object-oriented sense of the word ``class''.}
of statistical Model
which can be used independently or
combined to create structured Models.
Here mathematical transformations of statistical Models are added, and
the properties of transformations are considered and
they are implemented in the library;
this relies on defining and using certain subclasses of class Function.

MML is a Bayesian method of inference
devised by Wallace and Boulton \cite{WB68} in the 1960s,
their initial application being
mixture modelling (clustering, unsupervised classification \cite{JM08}).
MML was subsequently developed
both theoretically and practically and
has been used on many and varied problems \cite{Wal05}
including, but not limited to,
megalithic astronomical alignments \cite{PF88},
factor analysis \cite{WF92},
decision-trees \cite{WP93} and
protein structural alignments \cite{Col14}.
In general ``Strict'' MML inference is NP-hard \cite{FW02,Dow15}
but there are good and efficient approximations \cite{WF87,Wal05}
for many cases.
MML can be seen as a realisation of Ockham's razor \cite{Spa99,All18}.
The fact that it measures the complexity of statistical Models and of data
in the same units
makes it particularly suitable for
choosing between competing models and for
implementing structured Models.

MML inference \cite{WB68,Wal05} relies on
Bayes's theorem \cite{Bay63}\footnote{Typical usage is
   `$\pr(x)$' for $\pr(X=x)$,
   `$m$'  for a parameterised Model,
   `$sp$' for statistical parameters,
   `$upm$' for an unparameterised Model,
   `$ps$' for miscellaneous parameters,
   `$D$' for a data space,
   `$d \in D$' for a datum, and
   `$ds \in D^*$' for a data set of several data.}
\begin{equation}
\begin{split}
   \pr(m\,\&\,ds) & = \pr(m) \times \pr(ds|m) \\
                  & = \pr(ds) \times \pr(m|ds)
\end{split}
\end{equation}
and
on Shannon's mathematical theory of communication \cite{Sha48},
hence ``message'',
\begin{equation}
   I(E) = -\log(\pr(E))
\end{equation}
where $I(E)$ is the information content, or message length,
of an event~$E$.
Base-two logarithms measure information in `bits' and
natural  logarithms measure information in `nits' (natural bits).
Writing $msg(.)$ for $I(.)$ it follows that
\begin{equation}
   msg(m\,\&\,ds) = msg(m) + msg(ds|m)
\end{equation}
for Model (hypothesis) $m$ and data set~$ds$.
The message length of $m$ and of $ds$ together
is the length of a two-part message:
first transmit $m$ and then
transmit $ds$ assuming that $m$ is true.
Minimising the \textit{total} message length clarifies the trade-off
between Model complexity $msg(m)$ and its fit to the data $msg(ds|m)$;
the Model that achieves the minimum is the best Model,
the best answer to the inference problem being posed.
(A one-part message may be shorter,
although by a surprisingly small amount,
but it does not provide an answer to any inference problem.)
Note that MML considers the accuracy of measurement of
continuous data (\mysec{sec:Continuous}) and
the optimal precision to which parameters should be stated
so every continuous datum, and parameter, has a probability
not just a probability density.
This is one of the reasons that, in general,
MML is not the same as maximum a posteriori (MAP) estimation.
Even a discrete valued parameter of a Model may have
an optimal precision that is
less than its data type would allow.


\begin{figure}
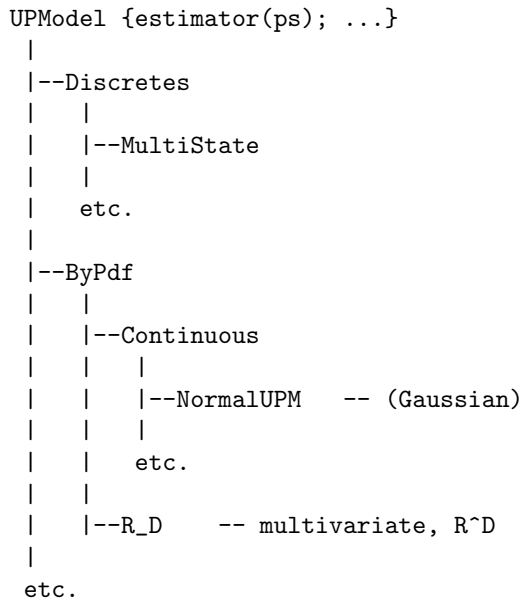

\begin{verbatim}
  UPModel {estimator(ps); ...}
   |
   |--Discretes
   |   |
   |   |--MultiState
   |   |
   |   etc.
   |
   |--ByPdf
   |   |
   |   |--Continuous
   |   |   |
   |   |   |--NormalUPM   -- (Gaussian)
   |   |   |
   |   |   etc.
   |   |
   |   |--R_D    -- multivariate, R^D
   |
   etc.
\end{verbatim}
\caption{Main UnParameterised Model classes}
\label{fig:UPModel}
\end{figure}

It is not the place of this paper to argue for
the usefulness of transformations of probability distributions or for
a certain transformed distribution being a good fit for a
particular kind of data -- these have been well-studied by statisticians.
Rather it examines a way of implementing such transformations
more expressively in programming languages.

\section{The Kinds of Model and Associates}
\label{sec:Models}

There are two stages of Model --
unparameterised Models which are instances of
class \refclass{mml}{UPModel}{UPModel}\footnote{
   Note, on a suitable device, but probably not in an email reader,
   hyperlinks in this pdf file
   such as \refclass{mml}{UPModel}{UPModel}$\leftarrow$(click)
   lead to online documentation and software within
   \myurl{https://www.cantab.net/users/mmlist/MML/A/} }
and parameterised Models which are instances of
class \refclass{mml}{Model}{Model}.

An unparameterised Model $upm$
can be applied to appropriate statistical parameter(s) $sp$
to create a parameterised Model $m=upm(sp)$,
for example,
$\refmember{mml}{MML}{N01}{N01}=\mathit{Normal}(\langle 0, 1\rangle)$
is the Normal Model (Gaussian distribution) with
mean $0$ and standard deviation~$1$.
An unparameterised Model has problem-defining parameters,
for example, \refclass{mml}{Discretes.Bounded}{Bounded}
has the bounds of its data space.
Problem-defining parameters are \textit{given}, not estimated.
For some, such as the Normal distribution,
the problem-defining parameters are trivial, $triv$ or $(\,)$, and
in such cases a single \refmember{mml}{MML}{Normal}{instance}
of the unparameterised Model is sufficient.
An unparameterised Model    (\myfig{fig:UPModel}) can
create parameterised Models (\myfig{fig:Model}).
Hopefully it can also
\refmember{mml}{UPModel}{estimator-la.la.Value-}{estimate}
(\myfig{fig:Estimator})
a parameterised Model to fit a
given data set $ds$.

\begin{figure}
\begin{verbatim}
  Model {pr(d); nlPr(d); ...}
   |
   |--Discretes.M   -- e.g. "fair coin"
   |   |
   |   etc.
   |
   |--ByPdf.M {pdf(d); ...}
   |   |
   |   |--Continuous.M
   |   |   |
   |   |   etc.
   |   |
   |   |--R_D.M    -- multivariate, R^D
   |
   etc.
\end{verbatim}
\caption{Main (Parameterised) Model classes}
\label{fig:Model}
\end{figure}

A parameterised Model has statistical parameters,
for example, the (standard) Normal has its mean and standard deviation.
In some cases
statistical parameters are trivial as, for example, in a
\refclass{mml}{Discretes.Uniform.M}{Uniform} distribution,
but
statistical parameters are generally \textit{estimated} from a given data set
although they can also be given, as with $N01$.
In accord with the MML framework \cite{WB68, Wal05},
an estimated Model has a message length, $msg_1$, and
the data set from which it was estimated has a message length, $msg_2$,
calculated under the assumption that the Model is true.
An MML estimator attempts to find a Model to
minimise the two-part message length,
$msg = msg_1 + msg_2$.
(In the case of \textit{given} statistical parameters, $msg_1$ is
zero as the parameters are common knowledge.)

The principal responsibility of a parameterised Model $m$ is to
give the probability $\pr(d)$ and
negative log probability~\footnote{
   Many calculations in the implementation are
   actually done in terms of negative log probabilities
   as those quantities are typically more manageable than
   plain probabilities
   (similar considerations may apply to probability densities).}
$nlPr(d)$
of a datum $d$ from $m$'s data space.
It may also do other things such as generate a random value,
\refmember{mml}{Model}{random--}{$random()$},
from its probability distribution.

\begin{figure}
\begin{verbatim}
  Estimator
    { ds2Model(ds);  -- data set -> Model
    ...}
\end{verbatim}
\caption{Estimator class}
\label{fig:Estimator}
\end{figure}

An \refclass{mml}{Estimator}{Estimator} (\myfig{fig:Estimator})
may have parameters to control its actions, for example,
to set the amount of lookahead in a search algorithm, or
to set a prior distribution
on the statistical parameters of the Models that it will estimate.

A further important class in the software
is \refclass{la/la}{Function}{Function} (\myfig{fig:Function}).
The library includes a simple interpreter for
the $\lambda$-calculus \cite{Chu41} and
Functions can be defined by $\lambda$-expressions.
Functions can also be defined by
\refclass{la/la}{Function.Native}{Native} code, that is Java code.
Functions of Continuous data
\refclass{la/la}{Function.Cts2Cts}{Cts2Cts},
in mathematics $\textbf{R} \rightarrow \textbf{R}$,
and
\refclass{la/la}{Function.CtsD2CtsD}{CtsD2CtsD},
$\textbf{R}^D \rightarrow \textbf{R}^D$,
will be important later.
Note that a UPModel is a Function because it can be
applied to statistical parameters to return a parameterised Model and
an Estimator is a Function because it can be
applied to a data set to return a parameterised Model.
Models, Functions, and hence UPModels and Estimators, are first class Values.

\begin{figure}
\begin{verbatim}
Function
 |  {apply(d); ...}
 |
 |--Lambda
 |
 |--Native
     |
     |--Cts2Cts {apply_x(x);
     |           d_dx(); ...} -- df/dx
     |
     |--CtsD2CtsD{J();        -- Jacobian
     |            nlJ(); ...} -- -log |J|
     |
     |--UPModel
     |
     |--Estimator
     |
     etc.

interface HasInverse{inverse();}
\end{verbatim}
\caption{Main Function classes}
\label{fig:Function}
\end{figure}


\section{Model Transformations}
\label{sec:Transformations}

Perhaps the most widely known transformed Model is
the log-Normal probability distribution:
for a data set
$ds = [d_1, d_2, ...]$,
$d_i \in (0, \infty)$,
it assumes that the values $[\log(d_1), \log(d_2), ...]$
are modelled by a Normal distribution.
Conversely, to generate a random value from a log-Normal,
generate a random value $x$ from the underlying Normal distribution and
apply $log^{-1}$ to $x$,
that is, return~$e^x$.
We will see more of the log-Normal in \mysec{sec:Continuous}.

\begin{figure}
\centering
\includegraphics[width=0.53\textwidth]{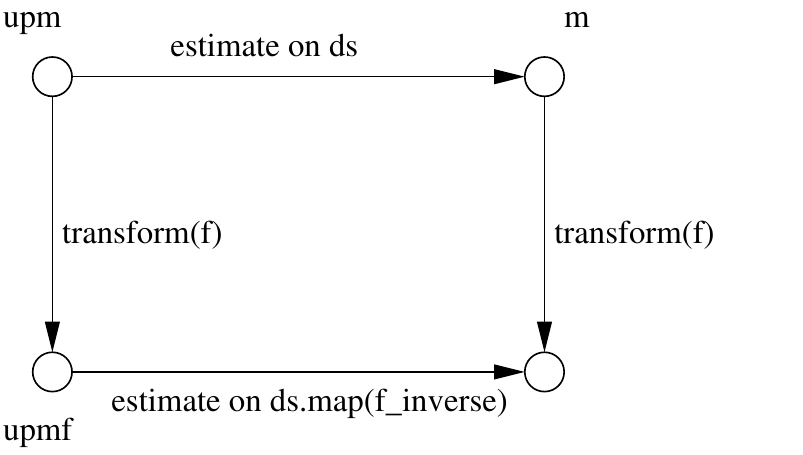}
\caption{Transforming and estimating}
\label{fig:tr-est}
\end{figure}

In general, a Model can be transformed by a one-to-one function, $f$,
having an inverse, $f^{-1}$.
Both unparameterised and parameterised Models can be transformed;
$\myMathit{upmf =} $\refclass{mml}{UPModel.Transform}{$upm.transform(f)$}
remains an unparameterised Model and
$\myMathit{mf =} $\refclass{mml}{Model.Transform}{$m.transform(f)$}
is a parameterised Model.
We also have that, \textit{as distributions},
transforming with $f$ and parameterising with $sp$ commute
\begin{equation}
   upm(sp).transform(f) = upm.transform(f)(sp);
\label{eq:commute_sp_f}
\end{equation}
the left and right sides of equation~\ref{eq:commute_sp_f}
have different histories but
effect the same probability distribution.
Similarly, as distributions,
estimating (on appropriate data) and transforming commute (\myfig{fig:tr-est})
\begin{equation}
\begin{split}
   & upm.estimator(ps)(ds).transform(f)          \\
=~ & upm.transform(f).estimator(ps)(ds.map(f^{-1})).
\end{split}
\label{eq:commute_est_f}
\end{equation}
In addition,
\begin{equation}
\begin{split}
   & upm.estimator(ps)(ds).msg() \\
=~ & upm.transform(f)            \\
   & ~~~   .estimator(ps)(ds.map(f^{-1})).msg()
\end{split}
\label{eq:info}
\end{equation}
that is,
the amounts of information in $ds$ and in $ds.map(f^{-1})$ are the same.
Conditions
(\ref{eq:commute_sp_f}),
(\ref{eq:commute_est_f}) and
(\ref{eq:info})
are a kind of ``invariance'' of probability distributions (Models).
Of course a general transformation operation
of an arbitrary Model
by an arbitrary one-to-one Function (of the right kind)
cannot possibly know what are equivalent transformations, if any,
on the statistical parameters of the arbitrary Model.

Given a data set, $ds$,
$\myMathit{upmf}$'s estimator operates by applying $f$
to all members \refmember{la/maths}{Vector}{map-la.la.Function-}{($map(f)$)}
of $ds$ and giving the result to $upm$'s estimator.
Model $\myMathit{mf}$~generates a $random()$ value
by getting $m$ to generate one and applying $f^{-1}$ to it.
(It is a quirk of common usage that
when transforming a Model with Function $f$
one applies $f$ to data and $f^{-1}$ to
random values generated by the Model.)

Since the transforming function $f$ must be one-to-one,
in the case of a discrete data space and its Models,
such a function $f$ must either
permute the data space in some simple way or
set up a one to one correspondence with a same sized space;
continuous data spaces,
\mysec{sec:Continuous} and \mysec{sec:R_D},
are more interesting.


\section{Continuous Models}
\label{sec:Continuous}

First note that,
due to the properties of object-oriented programming,
an unparameterised `\refclass{mml}{Continuous}{Continuous}' Model --
one of continuous data -- is a subclass of UPModel.
Hence an instance of Continuous can be transformed
by treating it just like
any other UPModel (\mysec{sec:Transformations}),
however,
the transformed result is then just a UPModel and is
not an instance of the Continuous subclass.
If we want the
\refmember{mml}{Continuous}{transform-la.la.Function.Cts2Cts-}{transformed}
Continuous itself to also be an instance of Continuous,
a little more work is required.
In particular a $pdf(.)$ must be defined for
the parameterised transformed Continuous:
For example,
we would like log-Normal to be a Continuous not just a UPModel,
see \myfig{fig:UPModel}.

\begin{figure}
\centering
\includegraphics[width=0.48\textwidth]{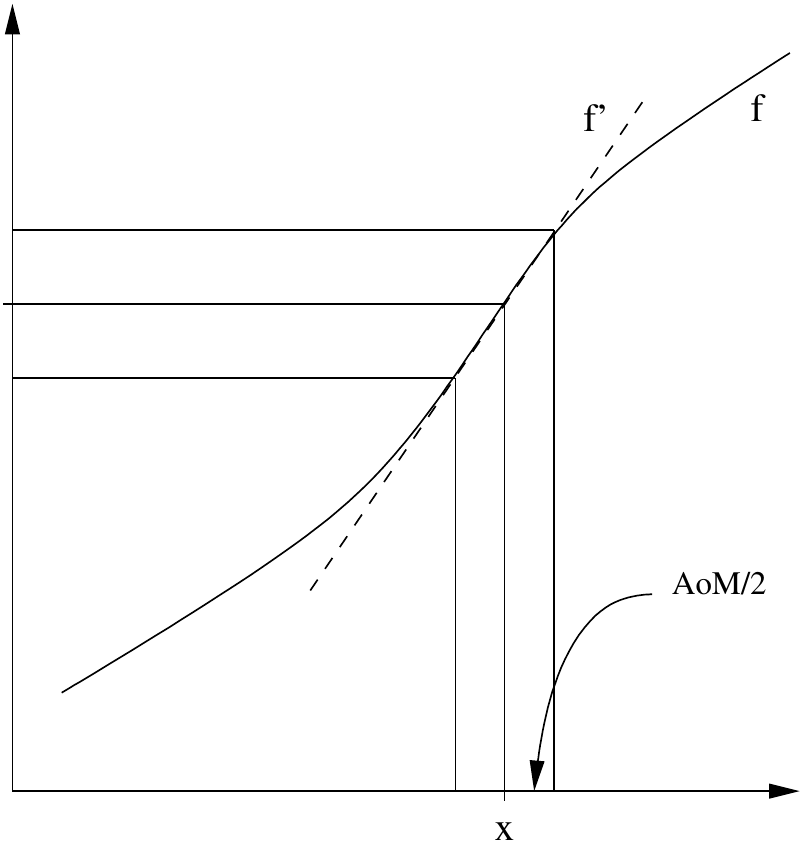}
\caption{AoM and Continuous}
\label{fig:cts}
\end{figure}

Each continuous datum, $d$, has a nominal value $x$ and
an accuracy of measurement
(\refmember{la/la}{Value.Cts}{AoM--}{$AoM$}) $\epsilon$ and
stands for $d = x \pm \frac{\epsilon}{2}$.
Usually it is assumed
that $\epsilon$ is small and
that a $pdf$ varies little across $x\pm \frac{\epsilon}{2}$
so that $\epsilon \times pdf(x)$ is a good approximation
for $\pr(x\pm\frac{\epsilon}{2})$.
When a \refclass{la/la}{Function.Cts2Cts}{continuous} function $f$
is applied to $d$
the result $f(d)$ has an AoM of $\epsilon \times |f'(x)|$
where
$f' = df/dx$ is the
\refmember{la/la}{Function.Cts2Cts}{d_dx--}{derivative}\footnote{Mathematically,
   most functions in $\textbf{R} \rightarrow \textbf{R}$
   are neither continuous nor differentiable
   but in practice most of those that we are \textit{interested} in
   are both continuous and differentiable over all or most of their domain.}
of $f$:
If the exact value of $d$ can be somewhere in a range of $\epsilon$,
$f(d)$ can be somewhere in a range of $\epsilon \times |f'(x)|$
(\myfig{fig:cts}).
For a continuous, one-to-one Function $f$ with an inverse $f^{-1}$
the \refmember{mml}{Continuous.Transform.M}{nlPdf_x-double-}{pdf}
of $m.transform(f)$ is
\begin{equation}
   m.pdf(f(x)) \times |f'(x)|.
\end{equation}
The factor $|f'(x)|$ adjusts the $pdf$ of $f(d)$ and,
in effect, adjusts the AoM of
$f(d)$ when $pdf(f(d))$ is used by $\pr(f(d))$.
Having a $pdf$ is sufficient to make the transformed Model an
instance of Continuous.

In the case of the
\refmember{mml}{MML}{logNormal}{log-Normal} Model,
            $f =$ \refmember{la/la}{Library}{log}{$log$},
$f$'s inverse    is  \refmember{la/la}{Library}{exp}{$exp$} and
$f$'s derivative is \refmember{la/la}{Library}{inv}{$1/x$}.
In the source code:
\begin{verbatim}
  logNormal = Normal.transform(log);
\end{verbatim}
Naturally
Function \refmember{la/la}{Library}{exp}{$exp$} has the
inverse    $log$  and
derivative $exp$, and
$transform(exp)$ turns a Model of $(0, \infty)$
into one of $(-\infty, \infty)$.


\section{$R^{D}$ Models}
\label{sec:R_D}

Multivariate continuous data are members of $\textbf{R}^D$
for some dimension, $D$, and
an unparameterised Model of such data is an
instances of class
\refclass{mml}{R\_D}{R\_D}.\footnote{No(?) programming language
   allows R\string^D ($R^D$) as an identifier;
   R\_D is the closest we can get to it in program code.}
Note that each component of a measured multivariate datum
has an accuracy of measurement -- as in \mysec{sec:Continuous}.
As an example of transformation,
consider Cartesian coordinates in the plane
$\langle x, y \rangle \in \textbf{R}^2$
and polar coordinates,
$\langle r, \theta \rangle \in \textbf{R}_{+} \times [0, 2 \pi)
 \subset \textbf{R}^2$.
The functions
\refmember{la/la}{Library}{polar2cartesian}{$polar2cartesian$} and
\refmember{la/la}{Library}{cartesian2polar}{$cartesian2polar$}
effect mappings between these coordinate systems and
are inverses of each other.
If $upmc$ is an unparameterised Model of cartesian coordinates then
$upmp = upmc.transform(polar2cartesian)$ is a Model of polar coordinates.

When transforming a univariate Continuous Model with function $f$,
the derivative of $f$ was used to ``adjust'' the accuracy of measurement
of a datum.
With multivariate continuous data,
the \refmember{la/la}{Function.CtsD2CtsD}{J-la.maths.Vector-}{Jacobian} matrix
of $f$, and
its \refmember{la/la}{Function.CtsD2CtsD}{nlJ-la.maths.Vector-}{determinant},
take on that role.
A suitable $pdf(d)$ for $m.transform(f)$ is
\begin{equation}
   m.pdf(f(d)) \times |J(d)|.
\end{equation}

For polar2cartesian
\begin{equation}
J_{pc} =
\begin{pmatrix}
\cos \theta & -r \times \sin \theta \\
\sin \theta &  r \times \cos \theta
\end{pmatrix}
\end{equation}
and
for cartesian2polar
\begin{equation}
J_{cp} =
\begin{pmatrix}
 x/r   & y/r \\
-y/r^2 & x/r^2
\end{pmatrix}
\end{equation}
giving
$ |J_{pc}| = r $,
$ |J_{cp}| = 1/r $, and
$ J_{pc} \times J_{cp} = I$.


\subsection*{A Detail}
\label{sec:Detail}

The $pdf(.)$ of a transformed R\_D Model
calls upon the $pdf(.)$ of the Model being transformed and
the determinant of the Jacobian of the transforming Function.
This does not necessarily require the $AoM$
of each component of a transformed datum --
there is already provision for Vectors where the $AoM$ of a Vector as
a whole is known but not that of each component
(however every component of a \textit{measured} datum does have an AoM).
However some structured Models, such as
\refclass{mml}{Dependent}{Dependent} and
\refclass{mml}{Independent}{Independent},
apply sub-Models to one or more components (columns, variables) of data.
In such cases it may be be necessary
to attribute the $AoM$ of a transformed datum among its components.
This particularly arises in Estimators.
Therefore the
\refmember{la/la}{Function.CtsD2CtsD}{apply-la.la.Value-}{apply(.)} method of
a Function in $CtsD2CtsD$  ($\textbf{R}^D\rightarrow\textbf{R}^D$) uses
the \refmember{la/la}{Function.CtsD2CtsD}{J-la.maths.Vector-}{Jacobian}
matrix of the Function to set the ratios of
the result's component's $AoM$s and
uses its
\refmember{la/la}{Function.CtsD2CtsD}{nlJ-la.maths.Vector-}{determinant}
to scale them to arrive at the
correct total $AoM$ area (volume, \dots) in $\textbf{R}^D$.

The matter is relevant to Estimators because
the AoM of a datum influences the amount of information in the datum and
an Estimator trades-off
the complexity (information) of an estimated Model against
the complexity of a data set.
Other things being equal, doubling the AoM of a datum
reduces its information by one bit and,
in the limit, to know that $d = x \pm \infty$
is to know nothing at all about~$d$.
A sub-Model may need to know
how much information is in
those columns of the data in which it deals.


\section{Conclusion}
\label{sec:Conc}

Transformations have been implemented in the MML software library for
unparameterised \refclass{mml}{UPModel}{Models} and
parameterised   \refclass{mml}{Model}{Models} using
a one-to-one    \refclass{la/la}{Function}{Function} $f$.
To make the transformed Model's \refmember{mml}{Model}{random--}{random()} work
$f$ must have an inverse.
For Models of continuous data $f$'s derivative must be defined, and
for multivariate Models of continuous data $f$'s Jacobian must be defined.
As probability distributions,
parameterising and transforming a Model commute,
and
transforming and estimating (on corresponding data) commute.
Applying such a Function $f$ to all the members of a data set leaves the
information content of the data set unchanged.
In the source code the definition of the log-Normal distribution
is simply
\texttt{Normal.transform(log)} (\mysec{sec:Continuous}).

The author is not aware of any widely used programming language
where \textit{all} functions (subroutines, procedures, methods),
whether built-in (`$+$', `$-$', $sin$, $cos$ and so on) or user-defined,
are instances of an explicit `class Function'
(the consensus \cite{H2002} seems to be that it
would be possible in Haskell say).
Every function does actually have at least one method, `apply'.
Applying apply is almost invariably implicit -- $f\ x$ or $f(x)$ --
it is the space between the $f$ and the $x$,
and $(x)$ is just the same as $x$ after all.
Note that Haskell \cite{H98} also has an explicit alternative
(`\string$' as in $f\,\string$\,x$)
for apply.
Given a `class Function', subclasses and interfaces such as `$1-1$',
continuous, differentiable, invertible and so on
are possible and, as suggested above, interesting and useful.

%
%






\bibliographystyle{compj}

\bibliography{paper}

\end{document}